\documentclass[aps,jcp,twocolumn,superscriptaddress,floatfix]{revtex4-2}

\usepackage{graphicx}
\usepackage{amsmath,empheq}
\usepackage{amssymb}
\usepackage{ifthen}
\usepackage{booktabs}

\usepackage{graphicx}
\usepackage{dcolumn}
\usepackage{bm}
\usepackage{gensymb}

\newcommand{\F}[2]{{\cal F}_{{\rm #1}}\left[#2\right]}
\newcommand{\Fid}{\mathcal{F}_\mathrm{id}}
\newcommand{\Fhs}{\mathcal{F}_\mathrm{hs}}
\newcommand{\Fatt}{\mathcal{F}_\mathrm{att}}

\newcommand{\kB}{k_{\rm B}}
\newcommand{\imi}{\mathrm{i}}
\newcommand{\pln}{\mathrm{pl.}} 
\newcommand{\sph}{\mathrm{sph.}} 
\newcommand{\cyl}{\mathrm{cyl.}} 

\newcommand{\pdif}[2]{\ensuremath{\frac{\partial{#1}}{\partial{#2}}}}\newcommand{\pdiff}[3][]{\frac{\partial^{#1}#2}{\partial #3^{#1}}}

\newcommand{\rhob}{\ensuremath{\rho_\mathrm{b}}}
\newcommand{\sumq}{\sum_{t}}

\newcommand{\bb}{\begin{equation}}
\newcommand{\ee}{\end{equation}}
\newcommand{\ba}{\begin{eqnarray*}}
\newcommand{\ea}{\end{eqnarray*}}

\newcommand{\dd}{{\rm d}}
\newcommand{\rr}{{\mathbf r}}

\newcommand{\nn}{\mathbf{n}}
\newcommand{\oo}{\bm{\omega}}
\newcommand{\rc}{r_\mathrm{c}}
\newcommand{\eps}{\varepsilon}
\newcommand{\uatt}{u_\mathrm{att}}
\newcommand{\Uatt}{U_\mathrm{att}}
\newcommand{\nrm}[1]{\left|{#1}\right|}
\newcommand{\R}{\varrho}

\makeatletter
\def\@frameeq#1{%
  \framebox{$\,\displaystyle#1\hbox{\vrule height 2.4ex depth 1.4ex width 0pt}\,$}}
\newcommand\Equation[1]{$$\refstepcounter{equation}%
  \@frameeq{#1}%
  \eqno \hbox{\@eqnnum}$$\@ignoretrue\ignorespaces}
\newcommand\Displaystyle[1]{$$\@frameeq{#1}$$\@ignoretrue\ignorespaces}
\makeatother

\usepackage{longtable}
\usepackage[caption=false]{subfig}
\usepackage[allcolors=black,colorlinks]{hyperref}
 \usepackage{nameref}

\begin{document}

\title{Uniform distributions in nonuniform systems: Wall potentials generating constant density profiles in classical density functional theory}

\author{Jiří \surname{Janek}}
\email{janek@icpf.cas.cz}
\affiliation{Research group of Molecular and Mesoscopic Modelling, The Czech Academy of Sciences, Institute of Chemical Process Fundamentals, 165 02 Prague, Czech Republic}

\author{Alexandr \surname{Malijevský}}
\email{malijevsky@icpf.cas.cz}
\affiliation{Research group of Molecular and Mesoscopic Modelling, The Czech Academy of Sciences, Institute of Chemical Process Fundamentals, 165 02 Prague, Czech Republic}
\affiliation{Department of Physical Chemistry, University of Chemistry and Technology Prague, 166 28 Prague, Czech Republic}

\date{\today}

\begin{abstract}
We study the inverse problem of classical density functional theory for inhomogeneous fluids: finding the wall potential that produces a constant equilibrium density profile, i.e., a perfectly flat density distribution in the accessible region adjacent to a substrate. Within Rosenfeld's fundamental measure theory, we solve this problem for a one-component fluid in planar, spherical, and cylindrical geometries, considering both a hard-sphere fluid and a fluid with an additional truncated Lennard-Jones attraction treated at the mean-field level. Explicit analytical expressions are obtained for planar walls, while spherical walls also admit an analytical treatment in a more cumbersome form. The cylindrical case is treated numerically. The construction provides an explicit microscopic realization of structure-cancelling wall fields, related to flat-profile conditions that occur under special matching conditions in interfacial theories of wetting and drying. The theory also yields a compact collection of formulae for weighted densities and one-body direct correlation functions in the three fundamental geometries, providing useful reference expressions for density-functional implementations. The resulting analytic wall potentials are validated in independent density functional calculations, which confirm that the prescribed flat profiles are recovered within numerical accuracy.
\end{abstract}

\maketitle

\section{Introduction}

Classical Density Functional Theory (DFT) provides one of the most powerful microscopic frameworks for describing equilibrium structure of inhomogeneous fluids near substrates, interfaces, and confinements \cite{Evans1979, Henderson_Fundamentals, Hansen2006, Lutsko2010}. In its classical formulation, the equilibrium density profile follows from minimization of a grand-potential functional containing the intrinsic fluid free energy and the external potential exerted by the substrate. For simple fluids, modern nonlocal theories based on Rosenfeld's fundamental measure theory (FMT) \cite{Rosenfeld1989}  capture packing effects with high accuracy and have become standard tools for studying adsorption, capillary condensation, wetting, drying, and related interfacial phenomena in planar and curved geometries  \cite{Cuesta2002, Roth2010}.

Most applications of DFT address the direct problem, namely the
determination of the equilibrium density profile for a prescribed
external field. By contrast, the inverse problem of constructing an
external field that yields a prescribed density profile has been explored
only rarely. An early exactly solvable example is Percus's treatment of
hard rods in one dimension, where the field required to generate a given
density distribution can be obtained explicitly \cite{Percus1976}.
Closer to the present problem, Henderson considered planar external
fields producing step-function density profiles at wall-fluid interfaces
\cite{Henderson1991}, and Ditz and Roth later studied interactions
designed to generate constant density profiles in mixtures
\cite{Ditz2021}. These works are important precedents because they show
that profile-flattening fields and interactions are not merely formal
curiosities. In those studies, however, the required external fields were obtained in formal or numerical form. Henderson
addressed the planar wall problem, whereas Ditz and Roth considered a
related radial fixed-particle geometry and used the resulting
flat-profile interaction in a binary-mixture setting.

Here we consider a complementary inverse problem: the explicit
construction of wall potentials that suppress all wall-induced structure
within a microscopic nonlocal DFT. The central question is whether the compensating wall field can
be obtained analytically, and how it is modified by substrate geometry.
We show that this is possible to a remarkable extent but depends strongly on the geometry. For planar walls, the analytic results can be written
explicitly and are relatively compact. For spherical walls, the inverse
problem remains analytically tractable, but the resulting expressions
are considerably more involved. For cylindrical walls, the reduction of
the FMT kernels leads to complete elliptic integrals, and we therefore
treat this case numerically. Thus the main results are the explicit planar and spherical analytic results, together with the numerical cylindrical construction, covering the three fundamental substrate geometries.

At first sight, a flat density profile in the vicinity of a wall may
appear counterintuitive. Realistic substrates normally generate packing
oscillations, depletion layers, or adsorption, and these features are
precisely what nonlocal density functionals are designed to describe.
The wall potentials constructed here have the opposite role: they cancel,
in a controlled and explicit way, the inhomogeneities generated by
hard-core correlations and, when present, by attractive fluid-fluid
interactions. They therefore provide a direct microscopic measure of the
external field required to suppress interfacial structure. As a
secondary by-product, these potentials also provide useful tests for
numerical DFT calculations, since a correct implementation should recover
the prescribed flat profile up to discretization error.

Besides its intrinsic interest, this problem is connected to interfacial
phase behaviour. Recent work by Evans, Stewart, and Wilding provided a
unified description of hydrophilic and superhydrophobic surfaces in
terms of wetting and drying transitions \cite{Evans2019}, thereby
renewing interest in the role of wall-fluid interactions in controlling
interfacial structure. In particular, analytic studies of wetting with
long-ranged forces by Dietrich and Napi\'orkowski
\cite{Dietrich1991}, and more recently by Parry and Malijevsk\'y
\cite{Parry2023}, showed that, under special matching conditions
between wall-fluid and fluid-fluid interactions, the density profile at
a wall can become exactly flat. In the latter work, this flat-profile
condition was linked to maximally multicritical wetting and drying
transitions and to the structure of the corresponding surface phase
diagrams. The present work addresses a different but complementary
question: can such a cancellation be realized explicitly within a
microscopic nonlocal density functional, and how is the required field
modified by substrate curvature?

We consider two one-component model fluids. The first is a purely
repulsive hard-sphere fluid described by the original Rosenfeld FMT. The
second supplements the hard-core repulsion by a truncated
Lennard-Jones-like attraction treated at the mean-field level. For each
model we determine the wall potential that generates a constant density
profile in planar, spherical, and cylindrical geometries, yielding six
basic cases in total. The resulting formulae make explicit how the
compensating field depends on packing, attraction, and curvature. In
this sense, the work provides analytic structure-cancelling wall fields
rather than a purely numerical inversion procedure.

The remainder of the paper is organized as follows. We first summarize
the DFT framework, while the geometry-specific one-dimensional forms of
the weighted densities and direct-correlation-function contributions for
planar, spherical, and cylindrical symmetry are collected in
Appendix~\ref{app:geometry_kernels}. We then use the prescribed constant
profiles to construct the corresponding wall potentials for the
hard-sphere and truncated Lennard-Jones fluids. The planar analytic
expressions are given in Appendix~\ref{app:planar_explicit}, whereas 
 the lengthy expressions for spherical geometry are
provided in the Supplemental Material. The cylindrical case is analyzed
numerically. Finally, we validate the derived potentials in independent DFT calculations and discuss the physical implications of the resulting structure-cancelling fields.

\section{Density-functional formulation in planar, spherical, and cylindrical symmetry\label{sec:DFT}}

In this section we summarize the density-functional framework used throughout this work and formulate the geometry-specific expressions required for the subsequent analysis. For the planar, spherical, and cylindrical geometries considered here, the three-dimensional convolutions of Rosenfeld's fundamental measure theory can be reduced to one-dimensional integral expressions.

\subsection{General framework}

Within the grand-canonical formulation, the equilibrium density profile is obtained by minimizing the grand-potential functional
\begin{equation}
	\Omega[\rho] = \F{}{\rho} + \! \int \! \rho(\rr)\left[V(\rr)-\mu\right]\,\dd\rr,
	\label{eq:Omega}
\end{equation}
where $\F{}{\rho}$ is the intrinsic free-energy functional containing all interparticle interactions, $V(\rr)$ is the external potential generated by the substrate, and $\mu$ is the chemical potential of the bulk reservoir. For the one-component fluid considered here, we approximate the intrinsic free energy as
\begin{equation}
	\F{}{\rho} = \F{id}{\rho} + \F{hs}{\rho} + \F{att}{\rho},
	\label{eq:Fdecomp}
\end{equation}
where $\F{id}{\rho}$ is the ideal-gas contribution, $\F{hs}{\rho}$ describes the hard-sphere reference system, and $\F{att}{\rho}$ accounts for the attractive part of the intermolecular interaction.

The ideal-gas contribution is given exactly by
\begin{equation}
	\F{id}{\rho} = \beta^{-1} \! \int \! \rho(\rr)\left\{\ln\!\left[\Lambda^3\rho(\rr)\right]-1\right\}\,\dd\rr,
	\label{eq:Fid}
\end{equation}
where $\beta=1/\kB T$ is the inverse temperature and $\Lambda$ is the thermal de Broglie wavelength, usually set to unity.

Repulsive interactions are described by the original Rosenfeld fundamental measure theory (FMT) \cite{Rosenfeld1989},
\begin{equation}
	\F{hs}{\rho} = \beta^{-1} \! \int \!\Phi(\rr)\,\dd\rr,
	\label{eq:Fhs}
\end{equation}
with the reduced excess free-energy density $\Phi$ defined in terms of scalar and vector weighted densities $n_\alpha$ and $\nn_\alpha$ as
\begin{equation}
	\Phi = -n_0 \ln(1-n_3)
	+\frac{n_1 n_2 - \nn_1\cdot\nn_2}{1-n_3}
	+\frac{n_2^3 - 3\,n_2\,\nn_2\cdot\nn_2}{24\pi\left(1-n_3\right)^2}.
	\label{eq:Phi}
\end{equation}
The weighted densities are obtained by convolutions of the density profile with weight functions $\omega_\alpha$ and $\oo_\alpha$ corresponding to the fundamental geometric measures of hard spheres with diameter $\sigma$ and radius $R=\sigma/2$. Four of these weight functions are scalar,
\begin{equation}
	\begin{aligned}
		\omega_3(\rr) & = \Theta(R-\nrm{\rr}),  \\
		\omega_2(\rr) & = \delta(\nrm{\rr}-R),  \\
		\omega_1(\rr) & = \omega_2(\rr)/4\pi R, \\
		\omega_0(\rr) & = \omega_1(\rr)/R;      \\
	\end{aligned}
	\label{eq:omegascalardef}
\end{equation}
and two are vectorial,
\begin{equation}
	\begin{aligned}
		\oo_2(\rr) & = (\rr/\nrm{\rr})\,\delta(\nrm{\rr}-R), \\
		\oo_1(\rr) & = \oo_2(\rr)/4\pi R,
	\end{aligned}
	\label{eq:omegadef}
\end{equation}
where $\Theta$ is the Heaviside step function and $\delta$ denotes the Dirac delta distribution. The corresponding scalar weighted densities are
\begin{equation}
	n_\alpha(\rr) = \int\!\rho(\rr')\,\omega_\alpha(\rr-\rr')\,\dd\rr',
	\quad \alpha=0,1,2,3;
	\label{eq:scalarweightdensdef}
\end{equation}
while the vector weighted densities are
\begin{equation}
	\nn_\alpha(\rr) = \int\!\rho(\rr')\,\oo_\alpha(\rr-\rr')\,\dd\rr',
	\quad \alpha=1,2.
	\label{eq:vectorweightdensdef}
\end{equation}

Attractive interactions are treated at the mean-field level through
\begin{equation}
	\F{att}{\rho} = \frac{1}{2}\iint\!
	\rho(\rr)\rho(\rr')\uatt(\nrm{\rr'-\rr})\,\dd\rr\,\dd\rr',
	\label{eq:Fatt}
\end{equation}
where $\uatt$ is a radially symmetric attractive pair potential.

Minimization of the grand potential yields the Euler-Lagrange equation, which can be written in the self-consistent form
\begin{equation}
	\rho(\rr) = \Lambda^{-3} \exp\!\left[\beta\mu -\beta V(\rr) + c^{(1)}(\rr)\right],
	\label{eq:Euler-Lagrange}
\end{equation}
where the one-body direct correlation function is defined by the functional derivative
\begin{equation}
	c^{(1)}=-\beta\frac{\delta\mathcal{F}}{\delta\rho(\rr)}.
\end{equation}
It is convenient to decompose $c^{(1)}$ into a hard-sphere contribution and an attractive contribution:
\begin{equation}
	c^{(1)} =  c^{(1)}_{\rm hs} + c^{(1)}_{\rm att} = -\beta\frac{\delta\Fhs}{\delta\rho(\rr)} - \beta\frac{\delta\Fatt}{\delta\rho(\rr)}.
	\label{eq:cdecomp}
\end{equation}
From Eq.~\eqref{eq:Fatt} it follows that
\begin{equation}
	c^{(1)}_\mathrm{att}(\rr) = -\beta \! \int \! \rho(\rr')\,\uatt(\nrm{\rr-\rr'})\,\dd \rr',
	\label{eq:catt}
\end{equation}
whereas for the hard-sphere part we write
\begin{equation}
	c^{(1)}_{\rm hs}(\rr) = \sum_\alpha c^{(1)}_\alpha(\rr),
	\label{eq:chsdecomp}
\end{equation}
with the individual contributions given by convolutions of derivatives of $\Phi$ with respect to the weighted densities and the corresponding weight functions,
\begin{equation}
	c^{(1)}_\alpha(\rr) = -\int\! \pdif{\Phi}{n_\alpha}(\rr)\,
	\omega_\alpha(\rr'-\rr)\,\dd\rr'.
	\label{eq:c1alphadef}
\end{equation}

The required derivatives of $\Phi$, together with the geometry-specific
one-dimensional forms of the weighted densities and the corresponding
contributions to $c^{(1)}$ for planar, spherical and cylindrical geometries, are collected in Appendix~\ref{app:geometry_kernels}.

\subsection{Model fluid \label{subsec:model}}

We consider two one-component fluid models. The first is a purely repulsive hard-sphere fluid, for which the particles interact via
\begin{equation}
	u_\mathrm{hs}(r) =
	\begin{cases}
		\infty, & r \leq \sigma, \\
		0,      & r > \sigma \,.
	\end{cases}
\end{equation}

The second model complements the hard-sphere repulsion by a truncated Lennard-Jones-like attractive tail,
\begin{equation}
	u_\mathrm{att}(r) =
	\begin{cases}
		-4\varepsilon \left(\frac{\sigma}{r}\right)^6, & \sigma < r \leq \rc, \\
		0,                                          & r > \rc \,,
	\end{cases}
\end{equation}
so that the total pair potential reads
\begin{equation}
	u_\mathrm{LJ}(r) = u_\mathrm{hs}(r) + u_\mathrm{att}(r) =
	\begin{cases}
		\infty,                                     & r \leq \sigma, \\
		-4\varepsilon \left(\frac{\sigma}{r}\right)^6, & \sigma < r \leq \rc, \\
		0,                                          & r > \rc \,.
	\end{cases}
	\label{eq:pairpotential}
\end{equation}

Because the hard-sphere contribution is described by the original Rosenfeld functional, the corresponding bulk reference fluid obeys the Percus-Yevick equation of state \cite{Rosenfeld1989}. The bulk chemical potential is
\begin{multline}
	\beta\mu_\mathrm{hs} = \beta\pdiff{\Fid}{\rhob} + \beta\pdiff{\Fhs}{\rhob} = \\
	\ln(\rhob) + \frac{5\eta^3 - 13\eta^2 + 14\eta}{2(1-\eta)^3} - \ln(1-\eta),
	\label{eq:mu_hs}
\end{multline}
where $\eta = \pi \rhob \sigma^3/6$ is the packing fraction. The corresponding pressure is
\begin{equation}
	\beta p_\mathrm{hs} = \rhob \frac{1 + \eta + \eta^2}{(1-\eta)^3}.
	\label{eq:pressure_hs}
\end{equation}
Including attractive interactions, the bulk chemical potential and pressure become
\begin{equation}
	\beta\mu_\mathrm{LJ} = \beta\mu_\mathrm{hs} + \beta \pdiff{\Fatt}{\rhob} = \beta\mu_\mathrm{hs} + 32\beta\varepsilon \eta \left[\left(\frac{\sigma}{\rc}\right)^3 -1\right]
	\label{eq:mu_LJ}
\end{equation}
and
\begin{equation}
	\beta p_\mathrm{LJ} = \rhob \left[\frac{1 + \eta + \eta^2}{(1-\eta)^3} + 16 \beta\eps\eta \left(\frac{\sigma}{\rc}\right)^3 - 1\right].
	\label{eq:pressure_LJ}
\end{equation}

\subsection{Problem formulation \label{subsec:problem}}

The standard DFT problem is to determine the equilibrium density profile for a prescribed external potential by solving Eq.~\eqref{eq:Euler-Lagrange}. In the present work we consider the inverse problem searching an external wall potential that generates a constant density profile in the accessible fluid region. Specifically, we require
\begin{equation}
	\rho^\pln (z) =
	\begin{cases}
		\rhob, & z>R;    \\
		0,     & z\leq R
	\end{cases}
	\label{eq:profile-const}
\end{equation}
for planar geometry, and
\begin{equation}
	\rho^{\sph/\cyl} (r) =
	\begin{cases}
		\rhob, & r>\R + R;    \\
		0,     & r\leq \R + R
	\end{cases}
	\label{eq:profile-const-spherical}
\end{equation}
for spherical and cylindrical substrates, where $\R$ denotes the radius of the spherical or cylindrical wall. These radii must exceed the cutoff distance $\rc$ of the attractive interaction in order to avoid interactions across the origin.

For a prescribed profile, the required external potential follows directly from the Euler-Lagrange equation:
\begin{equation}
	\beta V(\xi)=\beta\mu+c^{(1)}(\xi)-\ln\!\left[\Lambda^3\rho(\xi)\right],
	\label{eq:extpot}
\end{equation}
where $\xi=z$ for planar symmetry and $\xi=r$ for spherical and cylindrical symmetry.

In Sec.~III we use the prescribed profiles defined above to construct
and discuss the corresponding compensating wall potentials. The explicit
analytic expressions entering this construction are collected outside
the main text: the geometry-dependent FMT kernels are given in
Appendix~\ref{app:geometry_kernels}, the planar analytic expressions in
Appendix~\ref{app:planar_explicit}, and the more cumbersome spherical
expressions in the Supplemental Material. The planar and spherical
cases are thus treated analytically, whereas the cylindrical case is
evaluated numerically because the reduced kernels involve complete
elliptic integrals. In the following, we take $\sigma$ as the unit of
length and $\varepsilon$ as the unit of energy; the cutoff is chosen as
$r_c=2.5\,\sigma$, as is common in simulation studies.

\section{Results\label{sec:results}}

\subsection{Planar geometry}
\label{sec:planar_results}

We begin with the planar geometry, which serves as the simplest case
and as a reference for the curved substrates considered below. For the
prescribed profile in Eq.~(\ref{eq:profile-const}), the one-dimensional
FMT expressions collected in Appendix~\ref{app:geometry_kernels} can
be evaluated analytically. The resulting weighted densities are
piecewise polynomial functions, with nontrivial structure confined to
a layer of one hard-sphere diameter next to the wall.

Substitution of these weighted densities into the Rosenfeld functional
derivative gives a closed analytical expression for the hard-sphere
one-body direct correlation function $c_{\rm hs}^{\rm pl}(z)$. The
mean-field attractive contribution $c_{\rm att}^{\rm pl}(z)$ can
also be evaluated analytically for the truncated Lennard-Jones tail.
The explicit formulae, including the individual hard-sphere
contributions, are collected in Appendix~\ref{app:planar_explicit}.
Thus, in planar geometry, the inverse problem does not require any
numerical inversion: once $c^{(1)}(z)$ is known, the wall potential
follows directly from Eq.~(\ref{eq:extpot}).

\begin{figure}
	\includegraphics[width=\columnwidth]{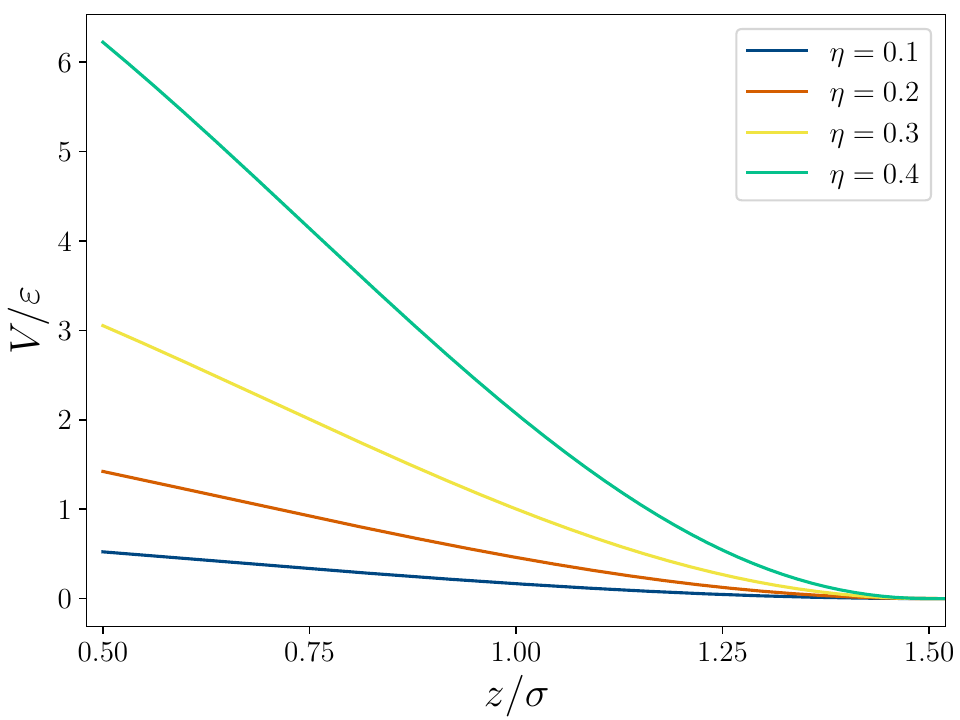}
	
	\vspace{0.2cm}
	\centerline{(a)}
	
	\vspace{0.35cm}
	\includegraphics[width=\columnwidth]{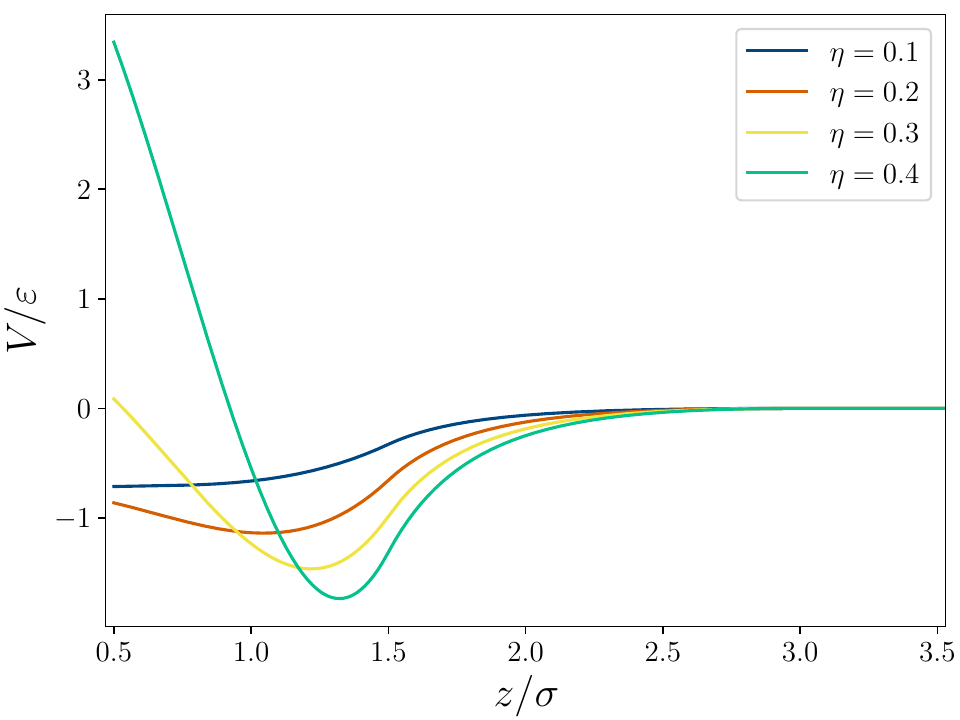}
	
	\vspace{0.2cm}
	\centerline{(b) }
	
	\caption{Planar wall potential  that generates the prescribed constant density profile for several packing fractions $\eta$. (a) Hard-sphere fluid.  (b)  Fluid with attractive interactions for  $T=1.5\,k_{\mathrm B}^{-1}\varepsilon$.}
	\label{fig:planar_potentials}
\end{figure}

\begin{figure}
	\includegraphics[width=\columnwidth]{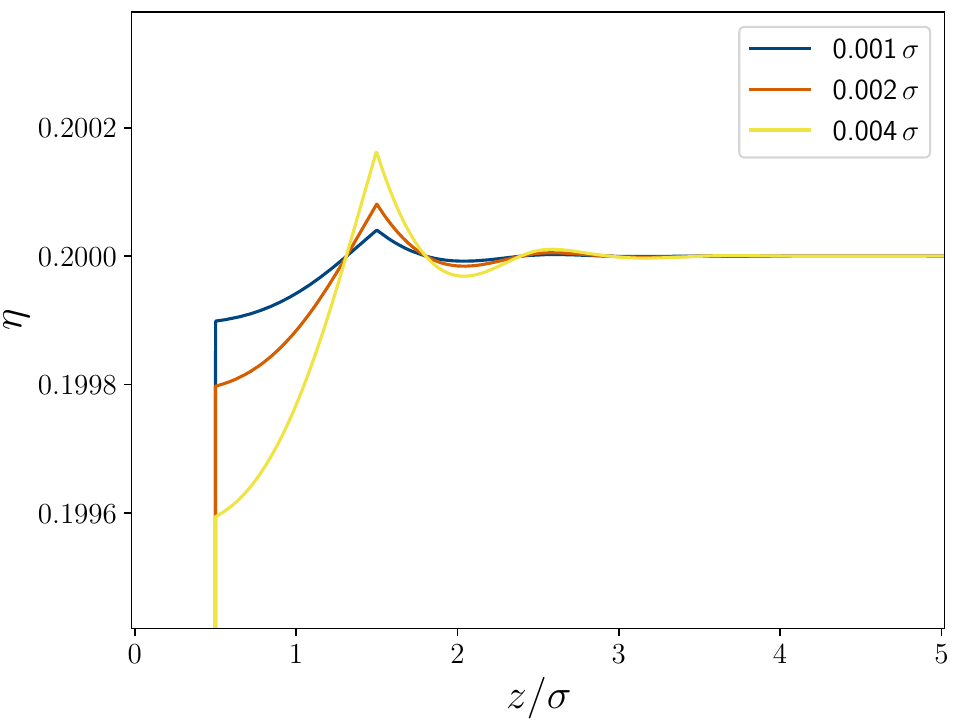}
	
	\vspace{0.2cm}
	\centerline{(a) }
		
	\vspace{0.35cm}
	\includegraphics[width=\columnwidth]{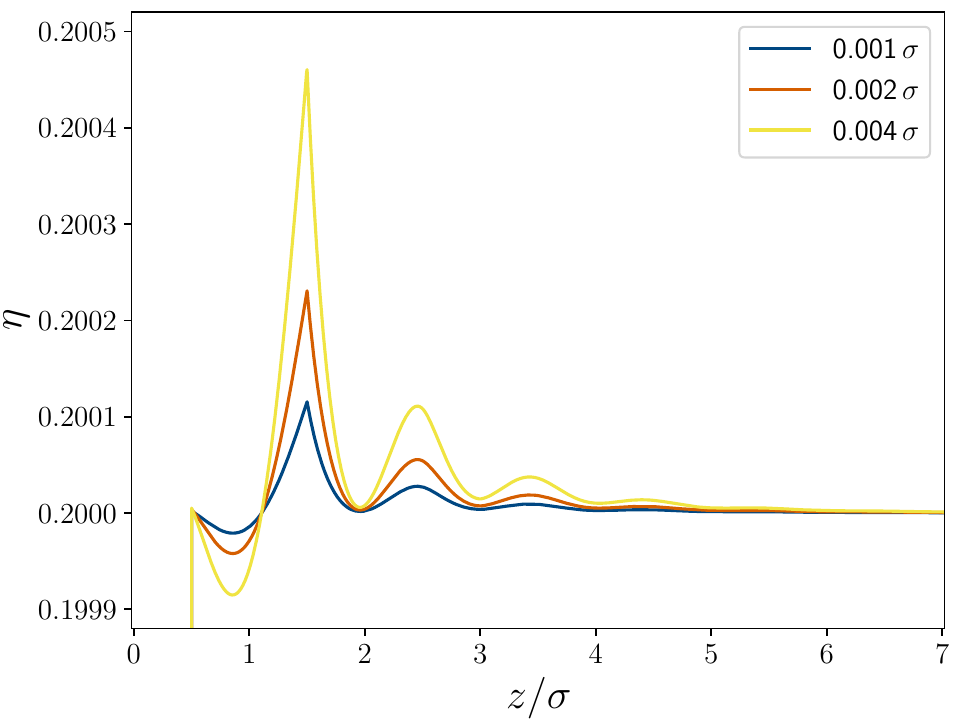}
	
	\vspace{0.2cm}
	\centerline{(b) }
	
\caption{DFT density profiles in planar geometry obtained with the analytically
	constructed external potentials. 
	(a) Hard-sphere fluid. 
	(b) Fluid with truncated Lennard-Jones attraction at
	$T=1.5\,\varepsilon/k_{\rm B}$.
	Several grid spacings are shown, demonstrating convergence toward the
	prescribed flat profile.}\label{fig:planar_validation}
\end{figure}

For the hard-sphere fluid, the resulting wall potential is shown in
Fig.~\ref{fig:planar_potentials}(a). It is purely repulsive and
short-ranged, with an amplitude that increases strongly with packing
fraction. The contact value obeys the exact sum rule
\begin{equation}
	\frac{\beta p_{\rm hs}}{\rho_b}
	=
	1+\beta V_{\rm hs}^{\rm pl}(R),
	\label{eq:contact_sum_rule_planar}
\end{equation}
which provides a useful consistency check on the analytical expression.

When the truncated attractive tail is included at the mean-field level,
the corresponding potential, shown in Fig.~\ref{fig:planar_potentials}(b),
has a broader range and develops an attractive part. This illustrates
the competition between the packing-induced hard-sphere contribution
and the cohesive mean-field attraction.

To verify the analytical wall potentials, we implemented them in
independent DFT calculations and computed the corresponding equilibrium
density profiles. The results are shown in
Fig.~\ref{fig:planar_validation}. In both the hard-sphere and
attractive cases, the prescribed constant profile is recovered within
numerical accuracy. The remaining deviations decrease systematically
upon refinement of the spatial grid, confirming that they originate
from discretization rather than from the analytical construction.

\subsection{Spherical geometry\label{subsec:results-spherical}}

We now turn to spherical symmetry and consider a solid sphere of radius $\varrho$
that generates the prescribed constant density profile (25). In contrast to the
planar case, curvature enters explicitly through the wall radius $\varrho$, so that the resulting expressions depend not only on the packing fraction but also on the substrate size. This allows us to assess how curvature modifies the compensating wall potential. As in the planar case, the prescribed profile can be substituted into the one-dimensional FMT expressions analytically. The resulting weighted densities are collected in the Supplementary Material, together with the mean-field attractive contribution.

The spherical weighted densities can be inserted into Eqs.~(A22)--(A27) to
obtain the hard-sphere contribution to the one-body direct correlation function.
In complete analogy with the planar case, the resulting expressions can be
worked out analytically. However, they are substantially more cumbersome, since
they depend  on both the packing fraction $\eta$ and the wall
radius $\varrho$  and are therefore deferred to the Supplemental Material. 

\begin{figure}
	\centering
	\includegraphics[width=\columnwidth]{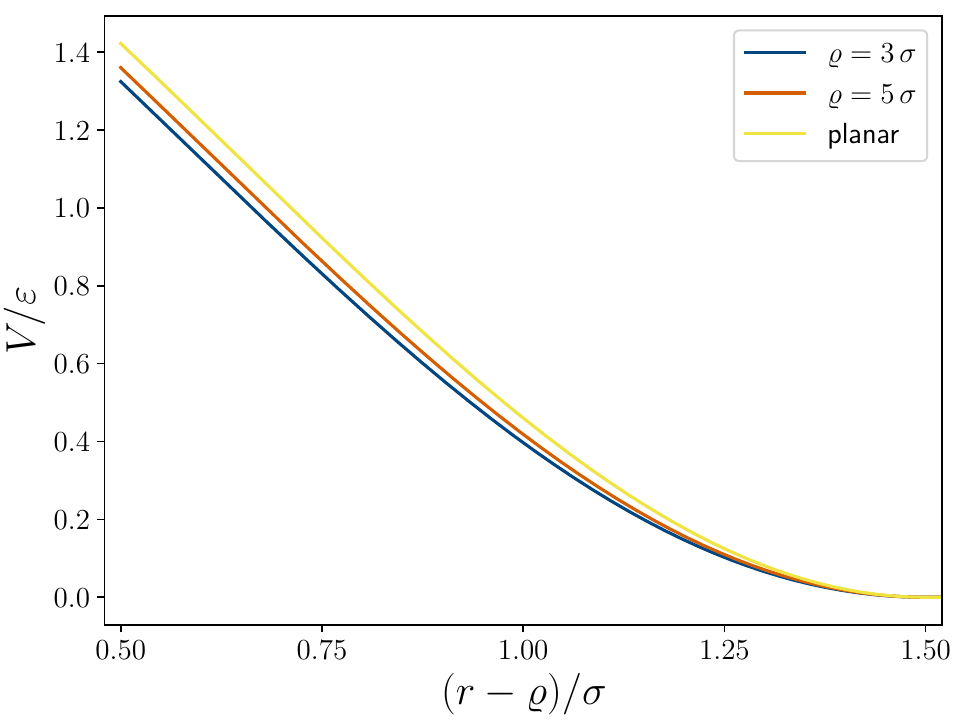}
	\caption{Hard-sphere wall potential in spherical geometry, $V_{\mathrm{HS}}^{\sph}(r-\varrho)$, for $\eta=0.2$. Results for two sphere radii are compared with the planar limit $V_{\mathrm{HS}}^{\pln}(z)$. The effect of curvature is relatively weak in the purely repulsive system.}
	\label{fig:extpot-spherical-HS}
\end{figure}

\begin{figure}
	\centering
	\includegraphics[width=\columnwidth]{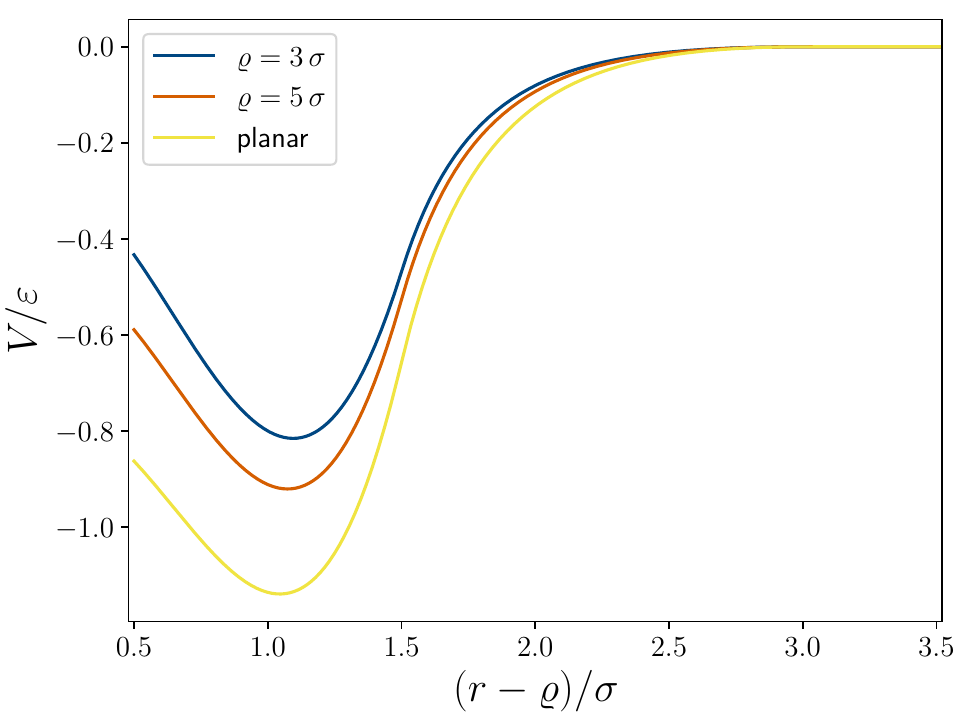}
	\caption{Wall potential in spherical geometry for the fluid with attractive interactions, $V_{\mathrm{LJ}}^{\sph}(r-\varrho)$, shown for several sphere radii and compared with the planar result. For $\eta=0.2$ and $T=1.5\,k_{\mathrm B}^{-1}\varepsilon$. In contrast to the hard-sphere case, curvature has a more pronounced effect on the depth and position of the attractive minimum.}
	\label{fig:extpot-spherical-LJ}
\end{figure}

Once the total spherical direct correlation function is known, the wall
potential follows directly from Eq.~(26). The resulting potentials for the
hard-sphere and attractive fluids are shown in Figs.~\ref{fig:extpot-spherical-HS}
and \ref{fig:extpot-spherical-LJ}, respectively, together with the planar limit for
comparison.  As expected, the spherical results converge to the corresponding planar results in the limit $\varrho\to\infty$.

For the hard-sphere fluid, the difference from the planar case is relatively
small for the radii considered here, indicating that the compensating field is
only weakly affected by curvature in the purely repulsive system. By contrast,
for the fluid with attractive interactions, the effect of curvature is more
pronounced: the attractive well becomes noticeably shallower for smaller
$\varrho$, showing that curvature affects the balance between packing and
mean-field attraction more strongly than in the purely repulsive hard-sphere
system.

\begin{figure}
	\includegraphics[width=\columnwidth]{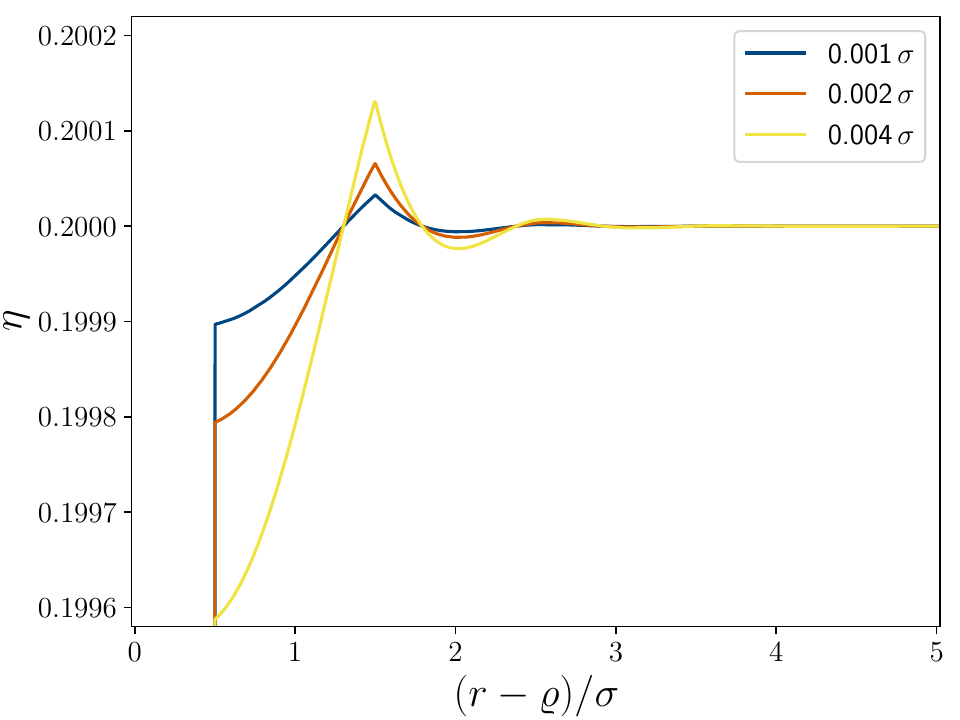}
	
	\vspace{0.2cm}
	\centerline{(a)}
	
	\vspace{0.35cm}
	\includegraphics[width=\columnwidth]{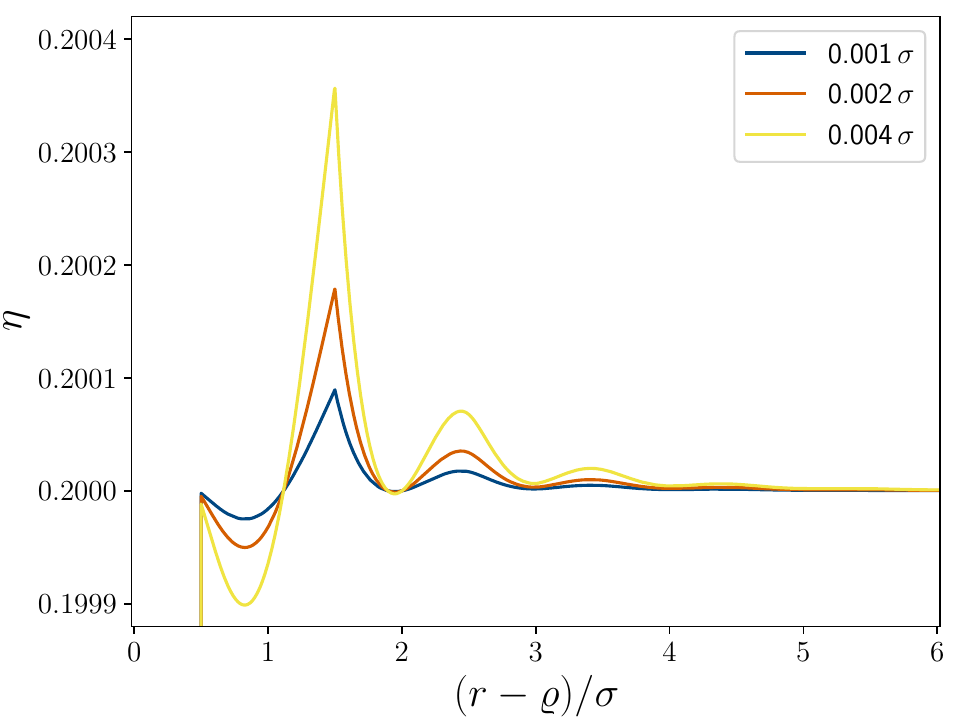}
	
	\vspace{0.2cm}
	\centerline{(b) }
	
	\caption{DFT density profiles in spherical geometry for $\eta_b=0.2$, $\varrho=3\sigma$, and $T=1.5\,k_{\mathrm B}^{-1}\varepsilon$, using the derived external potentials. (a) Hard-sphere fluid. (b) Fluid with attractive interactions. Several grid spacings are shown, demonstrating convergence toward the prescribed flat profile.}		
	\label{fig:densprof-sphere}
\end{figure}

To test the analytical construction, we implemented the spherical wall potentials
in independent DFT calculations and computed the corresponding equilibrium
density profiles numerically. The results are shown in
Fig.~\ref{fig:densprof-sphere}. The numerical profiles converge systematically
toward the prescribed constant profile as the grid spacing is reduced, confirming
that the analytically derived spherical wall potentials are implemented
correctly. For the finest grid, the maximal relative deviation is of order
$10^{-4}$, comparable to that found in the planar case.

\begin{figure}
	\includegraphics[width=\columnwidth]{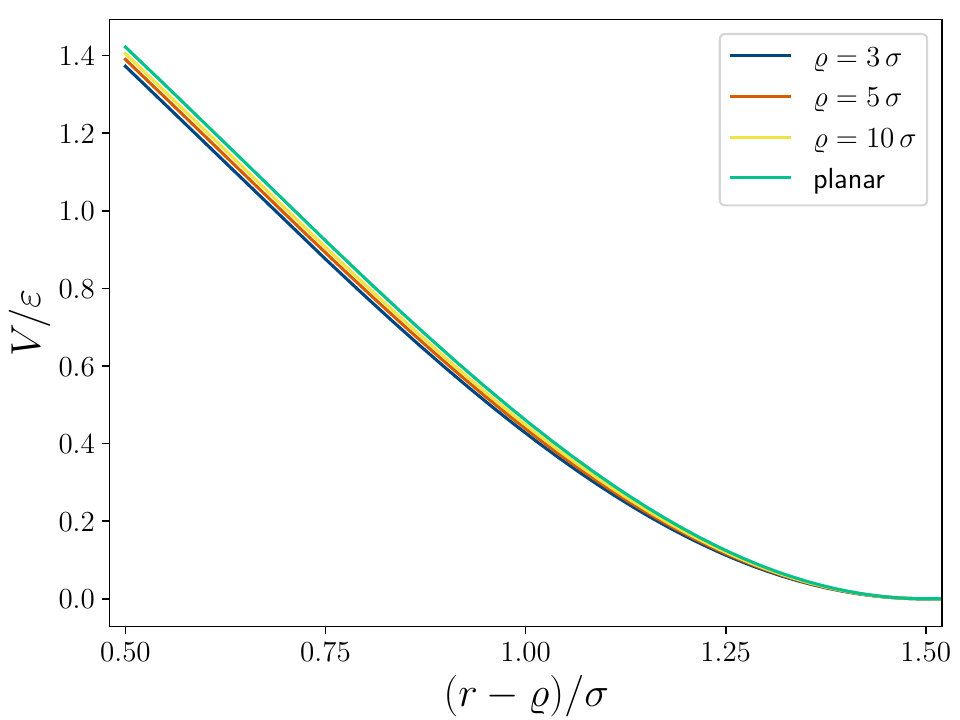}
	
	\vspace{0.2cm}
	\centerline{(a) }
	
	\vspace{0.35cm}
	\includegraphics[width=\columnwidth]{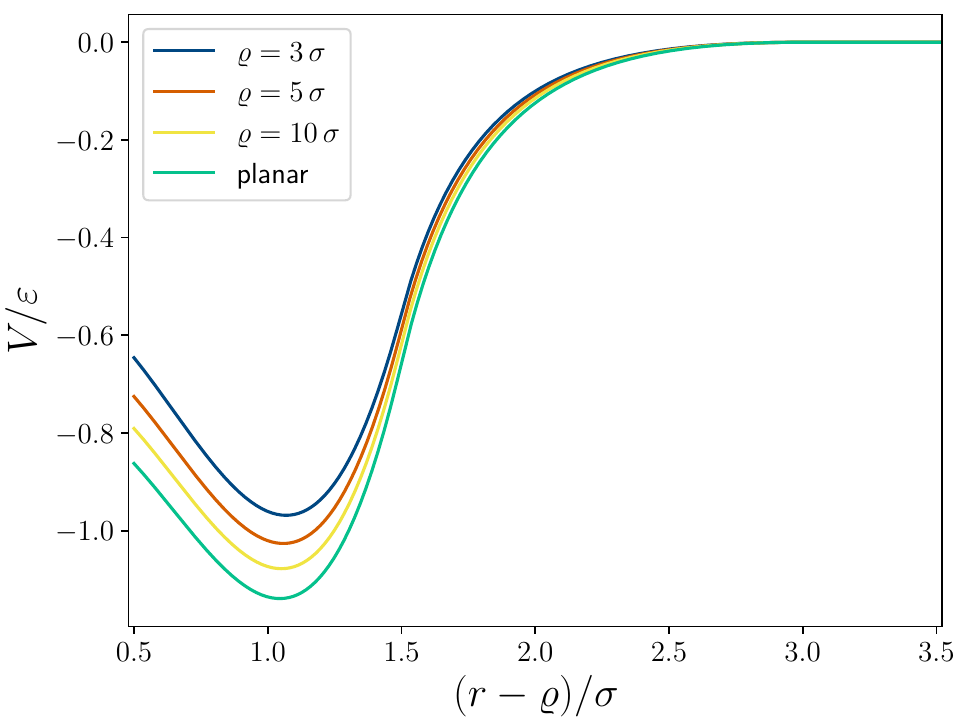}
	
	\vspace{0.2cm}
	\centerline{(b) }
	
	\caption{External wall potentials in cylindrical geometry obtained by numerical evaluation of the one-body direct correlation function, shown for several cylinder radii and compared with the planar result. (a) Hard-sphere fluid. (b) Fluid with attractive interactions at $T=1.5\,k_{\mathrm B}^{-1}\varepsilon$. In both cases the cylindrical potentials converge toward the planar limit as the cylinder radius increases. For $\eta=0.2$.}
	\label{fig:extpot-cyl}
\end{figure}

\begin{figure}
	\includegraphics[width=\columnwidth]{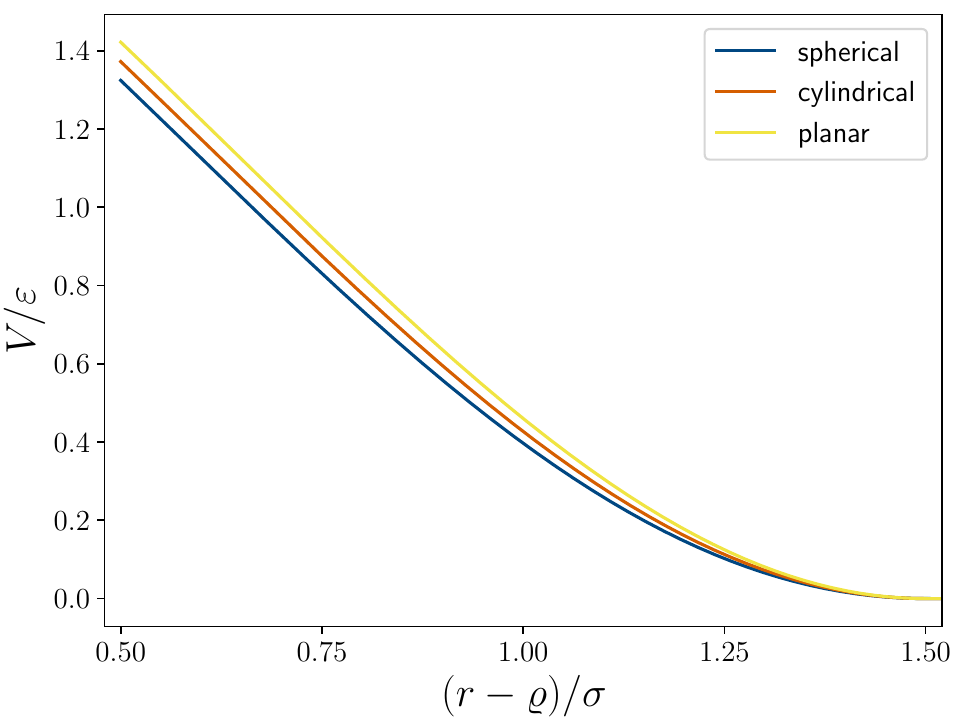}
	
	\vspace{0.2cm}
	\centerline{(a) }
	
	\vspace{0.35cm}
	\includegraphics[width=\columnwidth]{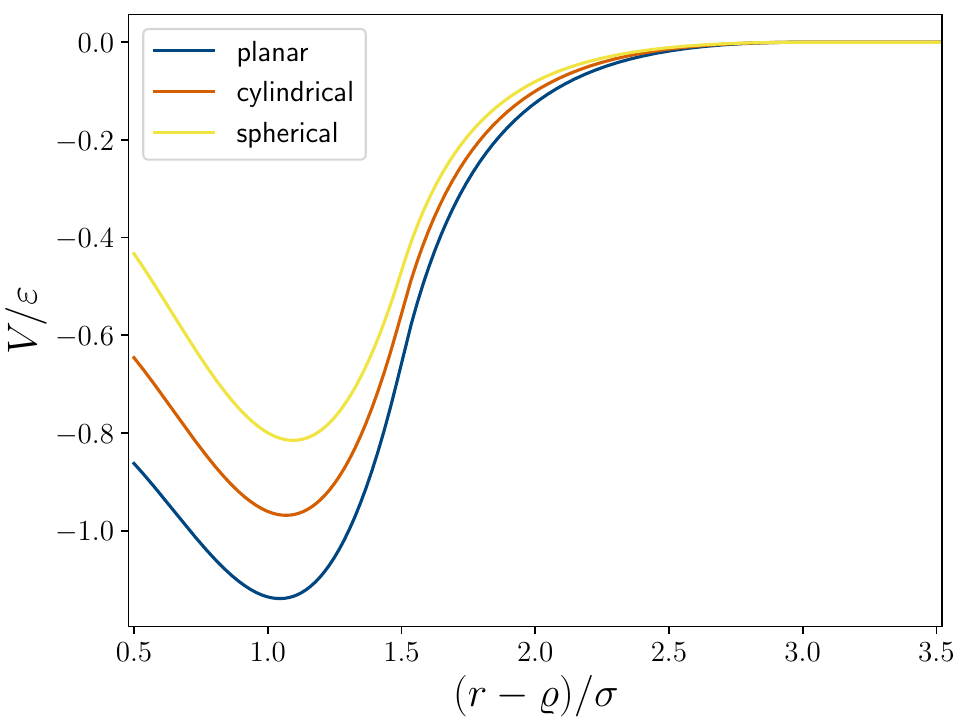}
	
	\vspace{0.2cm}
	\centerline{(b) }
	
	\caption{Comparison of wall potentials in planar, cylindrical, and spherical geometry for the same wall radius $\varrho=3\,\sigma$. (a) Hard-sphere fluid. (b) Fluid with attractive interactions at $T=1.5\,k_{\mathrm B}^{-1}\varepsilon$. For both model fluids, the cylindrical potential lies between the planar and spherical results, reflecting the smaller curvature of a cylinder relative to a sphere of the same radius. For $\eta=0.2$.}
	\label{fig:extpot-geom-compare}
\end{figure}

\subsection{Cylindrical geometry\label{subsec:results-cylindrical}}

We finally turn to the remaining fundamental geometry, namely an infinitely long solid cylinder of radius $\varrho$, for which the prescribed constant density profile is given by Eq.~\eqref{eq:profile-const-spherical}. Whereas the planar and spherical cases still admit explicit analytical treatment, the cylindrical reduction of the FMT kernels involves complete elliptic integrals and becomes substantially less tractable. Accordingly, the cylindrical case is treated numerically.

For the assumed cylindrical profile, we therefore evaluate the scalar and vector weighted densities, as well as all hard-sphere contributions to $c^{(1)}(r)$, by direct numerical quadrature of the one-dimensional integral
expressions shown in Appendix~\ref{app:geometry_kernels}. The attractive contribution is treated analogously by numerical evaluation of Eq.~\eqref{eq:catt_cylindrical}, so that the total cylindrical direct correlation function is obtained as the sum of the hard-sphere and attractive parts.

Having determined the cylindrical one-body direct correlation function numerically, the corresponding wall potential follows directly from Eq.~\eqref{eq:extpot}. Because the numerical procedure is based on the imposed flat density profile itself, the most informative results in this geometry are the resulting external potentials rather than a separate density-profile plot.

The cylindrical wall potentials for several cylinder radii are shown in Fig.~\ref{fig:extpot-cyl}, together with the planar result for comparison. In both the hard-sphere and attractive cases, the cylindrical potentials converge systematically toward the planar limit as $\varrho \to \infty$, as expected. For the hard-sphere fluid, the potential remains purely repulsive and differs only moderately from the planar case. For the fluid with attractive interactions, the cylindrical potential develops a pronounced minimum whose depth decreases as the cylinder radius decreases, reflecting the increasing effect of curvature.

Finally, Fig.~\ref{fig:extpot-geom-compare} compares the wall potentials for planar, cylindrical, and spherical symmetry for the same wall radius. In both model fluids, the cylindrical potential lies between the planar and spherical results and is approximately midway between them, reflecting the fact that cylindrical geometry represents an intermediate-curvature case.

\section{Conclusion}

We have studied the inverse problem of classical density functional theory for inhomogeneous fluids, namely the determination of external wall potentials that generate prescribed constant equilibrium density profiles. Employing Rosenfeld's fundamental measure theory for the hard-sphere reference system, we considered the three basic geometries of planar, spherical, and cylindrical symmetry. Two model fluids were investigated: a purely repulsive hard-sphere fluid and a fluid with an additional truncated Lennard-Jones attraction treated at a mean-field level.

Our analysis shows that this inverse problem admits an explicit solution to a remarkable extent. For planar walls, closed analytical expressions were obtained for the weighted densities, one-body direct correlation functions, and the resulting wall potentials.  For spherical walls, analytical results are also available, although the final expressions are substantially more involved. For cylindrical walls, the corresponding formulae appear not to allow for a closed form and were therefore evaluated numerically. 

The main value of these results is that the inverse construction can be
performed explicitly. The derived potentials show, in closed form for
planar and spherical substrates, how an external field must compensate
the packing and attractive contributions that would otherwise generate
interfacial structure. They also make transparent how this compensation
depends on density, intermolecular attraction, and substrate curvature,
information that would be less accessible from a purely numerical
inversion. As a useful by-product, the collected expressions and the
accompanying Fortran, C, and Python codes provide reference material for
FMT-based calculations and for checking numerical implementations. The
codes are available in the GitHub repository \cite{wall_potential_gh}.

In particular, while the planar and spherical cases share many
qualitative features, curvature modifies both the analytical structure
and the detailed shape of the wall potential. The inclusion of attractive
interactions changes the balance further and can qualitatively alter the
character of the resulting field compared with the purely repulsive
hard-sphere case. In this way, the present theory makes explicit how a
structureless density profile can arise in a nonuniform system through a
precise cancellation between substrate forces and fluid correlations.

The range of the compensating potential is tied to the
range of the kernels entering the density functional. For the
finite-ranged Rosenfeld-FMT plus mean-field attractive functional used
here, the constructed potential remains finite-ranged even at the
mean-field bulk critical point. Beyond the distance over which the FMT
weights and the truncated attractive fluid-fluid interaction overlap the
wall-induced density discontinuity, all weighted densities and attractive
convolutions take their bulk values. Equation~(\ref{eq:extpot}) then
reduces to the bulk Euler-Lagrange relation and gives $V=0$. This
conclusion should not be interpreted as a statement about the exact
critical fluid. Beyond mean-field theory, critical fluctuations generate
a diverging correlation length, and an exact compensating field could in
principle acquire long-ranged, ultimately algebraic, contributions at
criticality. Such fluctuation-induced critical effects lie outside the
present finite-ranged mean-field treatment.

The present construction also places the inverse problem in a broader
interfacial context. Flat-profile conditions encountered in wetting and
drying theories are not only asymptotic or phenomenological matching
conditions, but can also be realized explicitly within a microscopic
nonlocal density functional. In those interfacial theories, flat profiles
arise under special matching conditions between wall-fluid and
fluid-fluid interactions and are associated with maximally multicritical
surface behaviour. Here the same cancellation principle is implemented
directly: the external field is constructed so as to compensate the
packing and attractive contributions that would otherwise generate
interfacial structure. The resulting potentials therefore provide
explicit structure-cancelling wall fields and reveal how these fields
depend on substrate curvature.

Several extensions suggest themselves. One is to generalize the theory
to multicomponent fluids, where compensating fields may provide a route
to controlling adsorption selectivity or suppressing depletion-induced
structuring. It would also be useful to develop systematic curvature expansions
linking planar, spherical, and cylindrical results, and to assess whether
the cylindrical case admits compact semi-analytical approximations. On
the density-functional side, the construction could be repeated using
more refined hard-sphere functionals \cite{Tarazona2000, Roth2002, Malijevsky2006} or beyond-mean-field treatments of attractions. Finally, the same inverse viewpoint may be useful in dynamical settings \cite{Vrugt2020}, and possibly also in rheological contexts, where one could ask how transport, slip, or transient density relaxation are modified when the equilibrium wall-induced layering has been deliberately cancelled.

\section*{Supplementary Material}
The Supplementary Material contains the complete analytical expressions
for the spherical geometry, including the weighted densities, the
attractive contribution, and the hard-sphere one-body direct-correlation
contributions within Rosenfeld's fundamental measure theory.

\begin{acknowledgments}
We thank M. Posp\'\i\v sil for helpful discussions. The support of the Czech Science Foundation under project 25-15195S is gratefully acknowledged. 
\end{acknowledgments}

\appendix

\section{Geometry-dependent FMT kernels and one-body direct correlation functions}
\label{app:geometry_kernels}

In this appendix we collect the geometry-dependent one-dimensional
forms of the Rosenfeld FMT convolutions used in the main text.
These formulae define the weighted densities and the corresponding
hard-sphere contributions to the one-body direct correlation function
in planar, spherical, and cylindrical symmetry. They are used in Sec.~II
to construct the external potentials that generate the prescribed
constant density profiles.

\subsection{Derivatives of the Rosenfeld free-energy density}

We begin by listing the derivatives of the Rosenfeld excess
free-energy density $\Phi$ with respect to the scalar and vector
weighted densities. These derivatives enter the functional derivative
in Eq.~(\ref{eq:c1alphadef}).

\begin{equation}
	\begin{aligned}
		\pdif{\Phi}{n_0}   & = -\ln(1-n_3),                                                       \\
		\pdif{\Phi}{n_1}   & = \frac{n_2}{1-n_3},                                                 \\
		\pdif{\Phi}{n_2}   & =
		\frac{n_1}{1-n_3} + \frac{n_2^2-\nn_2\cdot\nn_2}{8\pi\left(1-n_3\right)^2},               \\
		\pdif{\Phi}{n_3}   & =
		\frac{n_0}{1-n_3}+\frac{n_1 n_2 - \nn_1\cdot\nn_2}{\left(1-n_3\right)^2}
		+ \frac{n_2^3-3\,n_2\,\nn_2\cdot\nn_2}{12\pi\left(1-n_3\right)^3},                        \\
		\pdif{\Phi}{\nn_1} & = -\frac{\nn_2}{1-n_3},                                              \\
		\pdif{\Phi}{\nn_2} & = -\frac{\nn_1}{1-n_3} - \frac{n_2 \nn_2}{4\pi\left(1-n_3\right)^2}.
	\end{aligned}
	\label{eq:Phi_derivatives}
\end{equation}

\subsection{Planar geometry}

We first consider planar symmetry, where the density profile depends
only on the coordinate $z$ normal to the substrate. After integrating
over the two directions parallel to the wall, the three-dimensional
FMT convolutions reduce to the following one-dimensional expressions.
We denote the nonzero components of the vector weighted densities by
$N^{\rm pl.}_{\alpha}(z)$:

\subsubsection{Weighted densities}

\begin{align}
	n^\pln_3(z) & = \pi \int_{z-R}^{z+R}\!\left[R^2 - \left(z-z'\right)^2\right]\rho(z')\,\dd z', \label{eq:n3planar} \\
	n^\pln_2(z) & = 2\pi R \int_{z-R}^{z+R}\!\rho(z')\,\dd z',                                    \label{eq:n2planar} \\
	n^\pln_1(z) & = \frac{1}{2}\int_{z-R}^{z+R}\!\rho(z')\,\dd z',                                \label{eq:n1planar} \\
	n^\pln_0(z) & =\frac{1}{2R}\int_{z-R}^{z+R}\!\rho(z')\,\dd z'.\label{eq:n0planar}\\
	N^\pln_2(z) & = 2\pi\int_{z-R}^{z+R}(z-z')\rho(z')\,\dd z',         \label{eq:N2planar} \\
	N^\pln_1(z) & = \frac{1}{2R}\int_{z-R}^{z+R}(z-z')\rho(z')\,\dd z'. \label{eq:N1planar}
\end{align}

\subsubsection{Contributions to the direct correlation function}

\begin{align}
	c_3^\pln(z)     & = -\pi \int_{z-R}^{z+R}\!\left[R^2 - \left(z-z'\right)^2\right] \pdif{\Phi^\pln}{n^\pln_3}(z')\,\dd z',
	\label{eq:c3planar}                                                                                                         \\
	c_2^\pln(z)     & = -2\pi R \int_{z-R}^{z+R}\! \pdif{\Phi^\pln}{n^\pln_2}(z')\,\dd z', \label{eq:c2planar}                        \\
	c_1^\pln(z)     & = -\frac 1 2 \int_{z-R}^{z+R}\! \pdif{\Phi^\pln}{n^\pln_1}(z')\,\dd z', \label{eq:c1planar}                     \\
	c_0^\pln(z)     & = -\frac 1 {2R} \int_{z-R}^{z+R}\! \pdif{\Phi^\pln}{n^\pln_0}(z')\,\dd z', \label{eq:c0planar}                  \\
	c_{N_2}^\pln(z) & = 2 \pi \int_{z-R}^{z+R}\! \left(z-z'\right) \pdif{\Phi^\pln}{N^\pln_2}(z')\,\dd z', \label{eq:cN2planar}       \\
	c_{N_1}^\pln(z) & = \frac 1 {2R} \int_{z-R}^{z+R}\! \left(z-z'\right)\pdif{\Phi^\pln}{N^\pln_1}(z')\,\dd z'. \label{eq:cN1planar}
\end{align}
The attractive contribution reduces to the one-dimensional convolution
\begin{equation}
	c^\pln_\mathrm{att}(z)  = -\beta \! \int_{-\infty}^{\infty} \! \rho(z')\,\Uatt(\nrm{z-z'})\,\dd z',
	\label{eq:cattplanar}
\end{equation}
where the effective planar kernel is obtained by integrating the pair attraction over the transverse coordinates,
\begin{equation}
	\Uatt(z) = \int_{-\infty}^{\infty}\!\int_{-\infty}^{\infty}\!\uatt\!\left(\sqrt{x^2+y^2+z^2}\right)\,\dd x\,\dd y.
	\label{eq:Uattplanar}
\end{equation}

\subsection{Spherical geometry}

For spherical symmetry, the density profile depends only on the radial
distance $r$ from the center of the spherical substrate. The angular
integrations can again be performed explicitly, reducing the FMT
convolutions to one-dimensional radial integrals \cite{Malijevsky2012}. The radial components of the vector weighted densities are denoted by
$N^{\rm sph.}_{\alpha}(r)$:

\subsubsection{Weighted densities}
\begin{align}
	n^\sph_3(r) & = \frac{\pi}{r} \int_{r-R}^{r+R} \! \rho(r') \, r'
	\left[R^2-\left(r-r'\right)^2\right]\, \dd r',                                  \label{eq:n3sph}\\
	n^\sph_2(r) & = \frac{2\,\pi R}{r} \int_{r-R}^{r+R} \! \rho(r') \, r' \,\dd r', \label{eq:n2sph}\\
	n^\sph_1(r) & =\frac{1}{2r} \int_{r-R}^{r+R} \! \rho(r') \, r' \,\dd r',        \label{eq:n1sph}\\
	n^\sph_0(r) & =\frac{1}{2Rr} \int_{r-R}^{r+R} \! \rho(r') \, r' \,\dd r',       \label{eq:n0sph}\\
	N^\sph_2(r) & = \frac{\pi}{r^2} \int_{r-R}^{r+R}\! \rho(r') \, r'
	\left(R^2+r^2-r'^2\right)\, \dd r',                                             \label{eq:N2sph}\\
	N^\sph_1(r) & = \frac{1}{4Rr^2} \int_{r-R}^{r+R}\! \rho(r') \, r'
	\left(R^2+r^2-r'^2\right)\,\dd r', \label{eq:N1sph}
\end{align}
where $N^\sph_\alpha(r)$ denotes the radial component of the corresponding vector weighted density.

\subsubsection{Contributions to the direct correlation function}
\begin{align}
	c_3^\sph(r)     & = -\frac{\pi}{r} \int_{r-R}^{r+R} \! \pdif{\Phi^\sph}{n^\sph_3}(r') r' \left[R^2-\left(r-r'\right)^2\right] \, \dd r', \label{eq:c3sph}\\
	c_2^\sph(r)     & = -\frac{2\,\pi R}{r} \int_{r-R}^{r+R} \! \pdif{\Phi^\sph}{n^\sph_2}(r') r' \,\dd r',                                 \label{eq:c2sph} \\
	c_1^\sph(r)     & = -\frac{1}{2r} \int_{r-R}^{r+R} \! \pdif{\Phi^\sph}{n^\sph_1}(r') r' \,\dd r',                                        \label{eq:c1sph} \\
	c_0^\sph(r)     & = -\frac{1}{2Rr} \int_{r-R}^{r+R} \! \pdif{\Phi^\sph}{n^\sph_0}(r') r' \,\dd r',                                       \label{eq:c0sph} \\
	c_{N_2}^\sph(r) & = \frac{\pi}{r^2} \int_{r-R}^{r+R}\!  \pdif{\Phi^\sph}{N^\sph_2}(r') r'\left(R^2+r^2-r'^2\right)\,\dd r',              \label{eq:cN2sph} \\
	c_{N_1}^\sph(r) & = \frac{1}{4Rr^2} \int_{r-R}^{r+R}\! \pdif{\Phi^\sph}{N^\sph_1}(r')r'\left(R^2+r^2-r'^2\right)\,\dd r'. \label{eq:cN1sph}
\end{align}
The attractive contribution becomes
\begin{equation}
	c^{\mathrm{sph.}}_{\rm att}(r) = -\frac{2\,\pi}{r} \beta \int_{-\infty}^{\infty}\int_{|r-r'|}^{\infty} \!  \rho(r')
	u_{\rm att}(r'') \, r''r'\,\dd r''\,\dd r'.
	\label{eq:catt_spherical}
\end{equation}

\subsection{Cylindrical geometry}

Finally, we consider cylindrical symmetry, where the density profile
depends only on the distance $r$ from the cylinder axis. In this case,
the reduction of the FMT convolutions leads to kernels involving complete
elliptic integrals. We use the standard notation
\begin{align}
	K(k) & = \int_0^{\pi/2}
	\frac{\dd\theta}{\sqrt{1-k^2\sin^2\theta}}, \\
	E(k) & = \int_0^{\pi/2}
	\sqrt{1-k^2\sin^2\theta}\,\dd\theta .
	\label{eq:ellipticintegralsdef}
\end{align}
With these definitions, the scalar and vector weighted densities, as well as the corresponding hard-sphere contributions to the one-body direct correlation function, take the following one-dimensional forms \cite{Malijevsky2007}.

\subsubsection{Weighted densities}

\begin{multline}
	n^\cyl_3(r)  = \\
	8 \int_{r-R}^{r+R}\! \rho(r') r' \sqrt{-a_2}
	\left[E\left(\sqrt{\frac{a_1}{a_2}}\right)-K\left(\sqrt{\frac{a_1}{a_2}}\right)\right]\,\dd r',
\end{multline}
\begin{equation}
	n^\cyl_2(r) = 8R \int_{r-R}^{r+R} \! \rho(r') r'
	\frac{K\left(\sqrt{a_1/a_2}\right)}{\sqrt{-a_2}}\,\dd r',
\end{equation}
\begin{equation}
	n^\cyl_1(r) = \frac{2}{\pi} \int_{r-R}^{r+R} \! \rho(r') r'
	\frac{K\left(\sqrt{a_1/a_2}\right)}{\sqrt{-a_2}}\,\dd r',
\end{equation}
\begin{equation}
	n^\cyl_0(r) = \frac{2}{\pi R} \int_{r-R}^{r+R} \! \rho(r') r'
	\frac{K\left(\sqrt{a_1/a_2}\right)}{\sqrt{-a_2}}\,\dd r',
\end{equation}
\begin{multline}
	N_2^\cyl(r)  = -\frac{4}{r}\int_{r-R}^{r+R}\!\rho(r') r'
	\left\{ \frac{r'^2-r^2-R^2}{\sqrt{-a_2}} K\left(\sqrt{\frac{a_1}{a_2}}\right)\right.\\
	\left.+ \sqrt{-a_2} \left[E\left(\sqrt{\frac{a_1}{a_2}}\right)-K\left(\sqrt{\frac{a_1}{a_2}}\right)\right] \right\}\,\dd r',
\end{multline}
\begin{multline}
	N_1^\cyl(r)  = -\frac{1}{\pi R r}\int_{r-R}^{r+R}\!\rho(r') r'
	\left\{ \frac{r'^2-r^2-R^2}{\sqrt{-a_2}} K\left(\sqrt{\frac{a_1}{a_2}}\right)\right.\\
	\left.+ \sqrt{-a_2} \left[E\left(\sqrt{\frac{a_1}{a_2}}\right)-K\left(\sqrt{\frac{a_1}{a_2}}\right)\right] \right\}\,\dd r',
\end{multline}
where
\begin{equation}
	\begin{aligned}
		a_1 & = R^2-\left(r'-r\right)^2, \\
		a_2 & = R^2-\left(r'+r\right)^2.
	\end{aligned}
	\label{eq:aidef}
\end{equation}

\subsubsection{Contributions to the direct correlation function}

\begin{multline}
	c^\cyl_3(r) = -8 \int_{r-R}^{r+R} \! \frac{\partial\Phi^\cyl}{\partial n^\cyl_3}(r') \,r' \sqrt{-a_2} \cdot \\
	\left[E\left(\sqrt{\frac{a_1}{a_2}}\right)-K\left(\sqrt{\frac{a_1}{a_2}}\right)\right]\,\dd r',
\end{multline}
\begin{equation}
	c^\cyl_2(r) = -8 R \int_{r-R}^{r+R} \! \frac{\partial\Phi^\cyl}{\partial n^\cyl_2}(r') \, r'
	\frac{K\left(\sqrt{a_1/a_2}\right)}{\sqrt{-a_2}}\,\dd r',
\end{equation}
\begin{equation}
	c^\cyl_1(r)  = -\frac{2}{\pi} \int_{r-R}^{r+R} \!\frac{\partial\Phi^\cyl}{\partial n^\cyl_1}(r') \, r'
	\frac{K\left(\sqrt{a_1/a_2}\right)}{\sqrt{-a_2}}\,\dd r',
\end{equation}
\begin{equation}
	c^\cyl_0(r) = -\frac{2}{\pi R} \int_{r-R}^{r+R} \! \frac{\partial\Phi^\cyl}{\partial n^\cyl_0}(r') \, r'
	\frac{K\left(\sqrt{a_1/a_2}\right)}{\sqrt{-a_2}} \,\dd r',
\end{equation}
\begin{multline}
	c^\cyl_{N_2}(r) = \\
	-\frac{4}{r}\int_{r-R}^{r+R} \!
	\frac{\partial\Phi^\cyl}{\partial N^\cyl_2}(r') \, r'
	\left\{ \frac{r'^2-r^2-R^2}{\sqrt{-a_2}} K\left(\sqrt{\frac{a_1}{a_2}}\right)\right.
	\\
	\left.+ \sqrt{-a_2} \left[E\left(\sqrt{\frac{a_1}{a_2}}\right)-K\left(\sqrt{\frac{a_1}{a_2}}\right)\right] \right\}\,\dd r',
\end{multline}
\begin{multline}
	c^\cyl_{N_1}(r) = \\
	-\frac{1}{\pi R r}\int_{r-R}^{r+R} \!
	\frac{\partial\Phi^\cyl}{\partial N^\cyl_1}(r') \, r'
	\left\{ \frac{r'^2-r^2-R^2}{\sqrt{-a_2}} K\left(\sqrt{\frac{a_1}{a_2}}\right)\right.
	\\
	\left.+ \sqrt{-a_2} \left[E\left(\sqrt{\frac{a_1}{a_2}}\right)-K\left(\sqrt{\frac{a_1}{a_2}}\right)\right] \right\}\,\dd r'.
\end{multline}
In these expressions, the derivatives of $\Phi^\cyl$ are obtained from
Eq.~(\ref{eq:Phi_derivatives}) by replacing the weighted densities by
their cylindrical counterparts. The attractive contribution is
\begin{multline}
	c^\cyl_\mathrm{att}(r) = \\
	8 \beta\int_\sigma^{\infty}\!\int_{r-r'}^{r+r'} \! u_{\rm att}(r')\, r'
	\rho(r'')\, r''
	\frac{K\left(\sqrt{b_1/b_2}\right)}{\sqrt{-b_2}}\,\dd r''\,\dd r',
	\label{eq:catt_cylindrical}
\end{multline}
where
\begin{align}
	b_1 & = r'^2-\left(r''-r'\right)^2, \\
	b_2 & = r'^2-\left(r''+r'\right)^2.
\end{align}

\vspace*{0.2cm}

\section{Planar analytic expressions for the prescribed flat profile}
\label{app:planar_explicit}

In this Appendix, all lengths are expressed in units of $\sigma$, so that $R=1/2$. 
Although the particle centers are accessible only for $z>R=1/2$, the weighted densities are nonzero already for $0<z<1/2$, because the FMT weight functions overlap with the accessible region. The weighted densities are listed over the full spatial range on which they are nonzero.

\subsection{Weighted densities}

Substituting the prescribed planar profile \eqref{eq:profile-const} into Eqs.~\eqref{eq:n3planar}--\eqref{eq:N1planar} yields simple closed expressions for all scalar and vector weighted densities:
\begin{align}
	n^\pln_3(z) & =
	\begin{cases}
		\eta \left(3z^2-2z^3\right), & 0<z<1, \\
		\eta,                        & z\geq 1, \\
		0,                           & \mathrm{otherwise},
	\end{cases} \\
	n^\pln_2(z) & =
	\begin{cases}
		6 \eta z,       & 0<z<1, \\
		6 \eta,         & z\geq 1, \\
		0,              & \mathrm{otherwise},
	\end{cases} \\
	n^\pln_1(z) & = \frac{n^\pln_2(z)}{2\pi}, \\
	n^\pln_0(z) & = 2n^\pln_1(z), \\
	N^\pln_2(z) & =
	\begin{cases}
		6\eta\left(z^2-z\right), & 0<z<1, \\
		0,                       & \mathrm{otherwise},
	\end{cases} \\
	N^\pln_1(z) & = \frac{N^\pln_2(z)}{2\pi}.
\end{align}

\subsection{Hard-sphere contribution to the one-body direct correlation function}

For completeness, we collect here the individual hard-sphere contributions entering the planar one-body direct correlation function,
\begin{equation}
	c_\mathrm{hs}^{\pln}(z)=c_3^{\pln}(z)+c_2^{\pln}(z)+c_1^{\pln}(z)+c_0^{\pln}(z)+c_{N_2}^{\pln}(z)+c_{N_1}^{\pln}(z).
\end{equation}
These expressions follow from Eqs.~\eqref{eq:c3planar}--\eqref{eq:cN1planar} after inserting the weighted densities for the prescribed planar profile. Since the external potential is required only in the accessible region,
we give the following expressions for $z\geq R=1/2$.

\begin{widetext}
	\begin{align}
		-c_3^\pln(z) & = \begin{cases}
			\dfrac{\eta\left(\eta^2 + \eta + 1\right)}{\left(1-\eta\right)^3}
			& z \geq \dfrac{3}{2}, \\[2ex]
			\sumq \Bigg\{
			\dfrac{1}{t\eta(1-\eta)^2(t-1)}
			\ln\!\left(\dfrac{2z-2t-1}{1-t}\right)
			\left[
			\dfrac{1}{2}
			+\left(\dfrac{1}{3}z^2+\dfrac{6t^2-10t+3}{6}z-\dfrac{8t+1}{12}\right)\eta^3
			\right.
			& \\[1ex]
			\qquad \left.
			+\left(\dfrac{18t-11}{24}z^2-\dfrac{24t^2-26t+15}{12}z+\dfrac{142t+59}{96}\right)\eta^2
			+\left(\dfrac{1}{2}z^2+\left\{t^2-2t\right\}z-\dfrac{t+3}{4}\right)\eta
			\right]
			\Bigg\}
			& \\[1ex]
			\qquad
			+\dfrac{32z^6-160z^5+224z^4+40z^3-294z^2+192z-37}
			{4(\eta-1)^3\left[2+(z-2)(2z-1)^2\eta\right]}\eta^4
			& \\[1ex]
			\qquad
			+\dfrac{128z^6-848z^5+2656z^4-4840z^3+4872z^2-2433z+558}
			{16(\eta-1)^3\left[2+(z-2)(2z-1)^2\eta\right]}\eta^3
			& \\[1ex]
			\qquad
			+\dfrac{128z^6-816z^5+1440z^4+104z^3-2256z^2+1929z-720}
			{16(\eta-1)^3\left[2+(z-2)(2z-1)^2\eta\right]}\eta^2
			& \\[1ex]
		    \qquad		
			+\dfrac{96z^4-400z^3+540z^2-396z+227}
			{8(\eta-1)^3\left[2+(z-2)(2z-1)^2\eta\right]}\eta
			& \\[1ex]
			\qquad
			+\dfrac{6z-9}{(\eta-1)^3\left[2+(z-2)(2z-1)^2\eta\right]}
			-3z\ln(2),
			& \dfrac{1}{2}\leq z < \dfrac{3}{2}, 
		\end{cases}
	\end{align}
	
\begin{align}
	-c_2^\pln(z) & =\begin{cases}
		\frac{3\eta\left(\eta+2\right)}{2\left(1-\eta\right)^2}{} & z \geq \frac{3}{2}           \\[1ex]
		\sum_t\frac{1+2 t }{8 t \left(1-t\right) \left(1-\eta \right)}\ln \! \left(\frac{2 z-2 t -1}{1-t}\right)+\frac{3 \eta  \left(2z -1\right) \left[\left(4z^{2}-8 z +1\right) \left(2z -1\right) \eta^{2}+\left(16 z^{3}-52 z^{2}+46 z -10\right) \eta -2z+11\right]}{8 \left(\eta-1 \right)^{2} \left[2+\left(z -2\right) \left(2z -1\right)^{2} \eta \right]}
		& \frac 1 2 \leq z < \frac 3 2, 
	\end{cases}
\end{align}%
\end{widetext}
\begin{align}
-c_1^\pln(z) & \nonumber                                                                                                                                                                                         \\
& \hspace{-3em}=\begin{cases}
	\frac{3\eta}{1-\eta}                                                                                           & z \geq \frac{3}{2}        \\[1ex]
	\sumq\frac{1}{2(1-t)}\ln \! \left(\frac{2 z-2 t -1}{1-t } \right) - \frac{3\left(2 z-1\right) \eta}{2(\eta-1)} & \frac 1 2 < z < \frac 3 2,
\end{cases}
\end{align}%

\begin{align}
-c_0^\pln(z) & \nonumber                                                                                                                                              \\
& \hspace{-3em}=\begin{cases}
	-\ln\left(1-\eta\right)                                                                               & z \geq \frac 3 2          \\[1ex]
	\sumq \frac{ \left(\eta  \,t^{2}-1\right)}{2 t \left(1-t \right) \eta}\ln \! \left(\frac{2 z -2 t-1}{1-t }\right)+\frac{9}{2}-3 z \\
	\quad + \left(z -\frac 1 2\right) \ln \! \left[(z-2)(2z-1)^2\eta+2\right]                             &                           \\[1ex]
	\quad -\left(z +\frac 1 2 \right) \ln \! \left(1-\eta \right)-\left(z -2\right) \ln \! \left(2\right) & \frac 1 2 < z < \frac 3 2,
\end{cases}
\end{align}%
\begin{align}
-c_{N_2}^\pln(z) & \nonumber                                                                                                                                                                                                                                        \\
& \hspace{-3em} =\begin{cases}
0& \hspace{-7em}z\geq\frac{3}{2}\\	
	\sumq \frac{1+\left[\left(2z -1\right) t^{2}-2\left(z +1\right) t +z\right] \eta}{2t \left(1-t\right) \eta}\ln \! \left(\frac{2 z-2 t -1 }{1-t }\right) &                                                                  \\[1ex]
	\quad +\frac{3 \left[\left(8z^{3}-22z^{2}+15z -3\right) \eta^{2}-\left(4z^{3}-12 z^{2}+9 z-5\right) \eta-2\right] \left(2z -3\right)}{2\left[2+\left(z -2\right) \left(2z -1\right)^{2} \eta \right] \left(\eta -1\right)} \\
	\quad + 3\left(z -\frac{1}{2}\right) \ln \! \left(2\right)
	& \hspace{-7em}\frac 1 2 \leq z < \frac 3 2,
\end{cases}
\end{align}%
\begin{align}
-c_{N_1}^{\mathrm{pl.}}(z) & \nonumber                                                                                                                                                                           \\
& \hspace{-3em} =\begin{cases}
0 & z\geq\frac{3}{2}\\                                         \\
	-\sumq \frac{1+\eta  \left(2z -1\right) t^{2}-2\eta  t z}{2\eta t \left(t -1\right)}\ln \! \left(\frac{2 z -2 t-1}{1-t}\right)   \\
	\quad+ 3z\left[1+\ln(2)\right]-\frac{3}{2}\left[3+\ln(2)\right]   & \frac 1 2 \leq z < \frac 3 2                                 
\end{cases}
\label{eq:SM-cN1planar}
\end{align}%
where all the sums over $t$ are over all three complex roots of the polynomial $P_3(t) = 2\eta t^3 - 3\eta t^2 + 1$ which are given by
\begin{equation}
\begin{aligned}
	t_1     & = \frac{\alpha}{2\eta}+\frac{\eta}{2\alpha}+\frac 1 2, \\
	t_{2,3} & = \beta \pm \imi \gamma,
\end{aligned}
\label{eq:SM-P3roots}
\end{equation}%
with $\alpha$, $\beta$, and $\gamma$ defined as
\begin{equation}
\begin{aligned}
	\alpha & = \sqrt[3]{\left(\eta - 2 + 2\sqrt{1-\eta}\right)\eta^2},                   \\
	\beta  & = -\frac{\alpha}{4\eta} - \frac{\eta}{4\alpha}+\frac{1}{2},                 \\
	\gamma & = \frac{\sqrt{3}}{2}\left(\frac{\alpha}{2\eta}-\frac{\eta}{2\alpha}\right),
\end{aligned}
\label{eq:SM-P3aux}
\end{equation}
and $\imi$ standing for the imaginary unit.
 The roots depend only on the packing fraction $\eta$ and are independent of $z$.

 The total correlation function, $	c_{\mathrm{hs}}^{\pln}(z)=\sum_\alpha c_\alpha^{\pln}(z)$, can be expressed as:
\begin{widetext}
	\begin{align}
		-c_\mathrm{hs}^\pln &=                                                                                                                                                                                                                                                                                                                   \nonumber                                                                                                                                                                                                                                       \\[1ex]
		& \hspace{-3em}=\begin{cases}
			\frac{5\eta^3 - 13\eta^2+14\eta}{2(1-\eta)^3} - \ln(1-\eta)                                                                                                                                                                                                                                                                                                       & \hspace{-7em}z\geq \frac 3 2                        \\[2ex]
			\sumq \Big\{ \frac{2z -1}{96\left(1-t\right) t \left(1-\eta \right)^{2}} \ln \! \left(\frac{2 z -2 t-1}{1-t}\right) \left[\left(48t^{2}-16 t -16 z -8\right) \eta^{2}-\left(96 t^{2}+36zt-70 t -22z-23\right) \eta +48t^{2}-24z+12\right]\!\Big\}                                                                                                                 &                                        \\[1ex]
			\qquad +\frac{64z^{6}-320 z^{5}+592 z^{4}- 544z^{3}+324 z^{2}-108 z +13}{8\left(\eta -1\right)^{3} \left[2+\left(2z -1\right)^{2} \left(z -2\right) \eta \right]} \eta^{4}   +\frac{128z^{6}-848 z^{5}+2176 z^{4}-2488 z^{3} +1032 z^{2}-141 z -36}{16\left(\eta -1\right)^{3} \left[2+\left(2z -1\right)^{2} \left(z -2\right) \eta \right]}\eta^{3}             &                                        \\[1ex]
			\qquad +\frac{128z^{6}-816 z^{5}+1824z^{4}-1864 z^{3}+1032 z^{2}-123z +90}{16\left(\eta -1\right)^{3} \left[2+\left(2z -1\right)^{2} \left(z -2\right) \eta \right]} \eta^2 + \frac{4z^{3}-12 z^{2}+3 z-5}{\left(\eta -1\right)^{3} \left[2+\left(2z -1\right)^{2} \left(z -2\right) \eta \right]}\eta - \left(z +\frac{1}{2}\right)  \ln \! \left(1-\eta \right) &                                        \\[1ex]
			\qquad +\left(z -\frac{1}{2}\right)  \left\{\ln \! \left[2+\left(2z-1\right)^2\left(z -2\right) \eta \right] +2\ln (2)\right\}
			& \hspace{-7em}\frac 1 2 < z < \frac 3 2 
		\end{cases}\label{eq:chs_planar}
	\end{align} 
\end{widetext}
where $\sumq$ denotes the sum over the roots of the cubic polynomial $P_3(t)$.

\subsection{Attractive contribution}
We now include the attractive contribution corresponding to the truncated Lennard-Jones-like tail, \eqref{eq:cattplanar}. For the planar target profile, the mean-field term can again be evaluated analytically:
\begin{multline}
	-c^\pln_\mathrm{att}(z)  = \beta \! \int_{z-\rc}^{z+\rc} \! \rho(z')\,\Uatt(\nrm{z-z'})\,\dd z' \\
	=\begin{cases}
		32\eta\beta\varepsilon \left(\frac{1}{\rc^3}-1\right),                                                               & z\geq \frac{1}{2}+\rc;         \\
		96\eta\beta\varepsilon \left(\frac{1-2\rc^3}{6\rc^3} + \frac{2z-1}{16\rc^4} + \frac{1}{3\left(2z-1\right)^3}\right), & \frac{3}{2}<z<\frac{1}{2}+\rc; \\
		48\eta\beta\varepsilon\left(\frac{1}{3\rc^3} -\frac{5}{24} + \frac{2z-1}{8\rc^4}- \frac{z}{4}\right),                & \frac{1}{2}<z\leq\frac{3}{2}.            \\
	\end{cases}
	\label{eq:catt_planar_explicit}
\end{multline}

\subsection{External wall potential}

Finally, using Eq.~(\ref{eq:extpot}), the  resulting wall potential  for planar geometry and the purely hard-sphere system is
\begin{equation}
\beta V_{\rm hs}^{\rm pl}(z)=
\beta\mu_{\rm hs}+c_{\rm hs}^{\rm pl}(z)-
\ln(\Lambda^3\rho_b),\qquad z\geq R,
\end{equation}
while for the fluid with the attractive tail, the potential reads
\begin{equation}
\beta V_{\rm LJ}^{\rm pl}(z)=
\beta\mu_{\rm LJ}+
c_{\rm hs}^{\rm pl}(z)+
c_{\rm att}^{\rm pl}(z)-
\ln(\Lambda^3\rho_b),\qquad z\geq R .
\end{equation}

\bibliography{wall_potential}

@Article{Rosenfeld1989,
  author    = {Rosenfeld, Yaakov},
  journal   = {Phys. Rev. Lett.},
  title     = {Free-energy model for the inhomogeneous hard-sphere fluid mixture and density-functional theory of freezing},
  year      = {1989},
  issn      = {0031-9007},
  month     = aug,
  number    = {9},
  pages     = {980--983},
  volume    = {63},
  comment   = {Rosenfeld's Fundamental Measure Theory (FMT)},
  doi       = {10.1103/physrevlett.63.980},
  publisher = {American Physical Society (APS)},
}

@Article{Malijevsky2012,
  author    = {Malijevský, Alexandr and Jackson, George},
  journal   = {J. Phys. Condens. Matter},
  title     = {A perspective on the interfacial properties of nanoscopic liquid drops},
  year      = {2012},
  issn      = {1361-648X},
  month     = oct,
  number    = {46},
  pages     = {464121},
  volume    = {24},
  comment   = {dependence of interfacial tension on curvature for small (comparable to particle diameter) radii
sign of the Tolman length as obtained from various approaches and methodologies
introduction offers quite a large review (9 pages)
comparison of various routes to surface tension calculation
expressions for c^(1) in spherical symmetry for Rosenfeld FMT},
  doi       = {10.1088/0953-8984/24/46/464121},
  publisher = {IOP Publishing},
}

@Article{Malijevsky2007,
  author   = {Malijevský, Alexandr},
  journal  = {J. Chem. Phys.},
  title    = {Fundamental measure theory in cylindrical geometry},
  year     = {2007},
  issn     = {0021-9606},
  month    = {04},
  number   = {13},
  pages    = {134710},
  volume   = {126},
  doi      = {10.1063/1.2713106},
}

@Misc{wall_potential_gh,
  author       = {Jiří Janek},
  howpublished = {GitHub repository},
  note         = {\url{https://github.com/janekj2727/wall_potential.git}},
  title        = {{wall\_potential}},
  year         = {2026},
  url          = {https://github.com/janekj2727/wall_potential.git},
}

@Article{Evans1979,
  author    = {Evans, R.},
  journal   = {Adv. Phys.},
  title     = {The nature of the liquid-vapour interface and other topics in the statistical mechanics of non-uniform, classical fluids},
  year      = {1979},
  issn      = {1460-6976},
  month     = apr,
  number    = {2},
  pages     = {143--200},
  volume    = {28},
  doi       = {10.1080/00018737900101365},
  publisher = {Informa UK Limited},
}

@Article{Percus1976,
  author    = {Percus, J. K.},
  journal   = {J. Stat. Phys.},
  title     = {Equilibrium state of a classical fluid of hard rods in an external field},
  year      = {1976},
  issn      = {1572-9613},
  month     = dec,
  number    = {6},
  pages     = {505--511},
  volume    = {15},
  doi       = {10.1007/bf01020803},
  publisher = {Springer Science and Business Media LLC},
}

@Article{Ditz2021,
  author    = {Ditz, Nikolas and Roth, Roland},
  journal   = {J. Chem. Phys.},
  title     = {Gas–liquid phase transition in a binary mixture with an interaction that creates constant density profiles},
  year      = {2021},
  issn      = {1089-7690},
  month     = may,
  number    = {20},
  volume    = {154},
  doi       = {10.1063/5.0048784},
  publisher = {AIP Publishing},
}

@Article{Evans2019,
  author    = {Evans, Robert and Stewart, Maria C. and Wilding, Nigel B.},
  journal   = {Proc. Natl. Acad. Sci.},
  title     = {A unified description of hydrophilic and superhydrophobic surfaces in terms of the wetting and drying transitions of liquids},
  year      = {2019},
  issn      = {1091-6490},
  month     = oct,
  number    = {48},
  pages     = {23901--23908},
  volume    = {116},
  doi       = {10.1073/pnas.1913587116},
  publisher = {Proceedings of the National Academy of Sciences},
}

@Article{Dietrich1991,
  author    = {Dietrich, S. and Napiórkowski, M.},
  journal   = {Phys. Rev. A},
  title     = {Analytic results for wetting transitions in the presence of van der Waals tails},
  year      = {1991},
  issn      = {1094-1622},
  month     = feb,
  number    = {4},
  pages     = {1861--1885},
  volume    = {43},
  doi       = {10.1103/physreva.43.1861},
  publisher = {American Physical Society (APS)},
}

@Article{Parry2023,
  author    = {Parry, Andrew O. and Malijevský, Alexandr},
  journal   = {Phys. Rev. Lett.},
  title     = {Surface Phase Diagrams for Wetting with Long-Ranged Forces},
  year      = {2023},
  issn      = {1079-7114},
  month     = sep,
  number    = {13},
  pages     = {136201},
  volume    = {131},
  doi       = {10.1103/physrevlett.131.136201},
  publisher = {American Physical Society (APS)},
}

@Book{Hansen2006,
  author    = {Hansen, Jean-Pierre and Ian R. McDonald},
  publisher = {Elsevier},
  title     = {Theory of simple liquids},
  year      = {2006},
  address   = {Amsterdam},
  edition   = {3rd edition},
  isbn      = {9780123870339},
  pagetotal = {619},
  ppn_gvk   = {1658035348},
}

@Book{Henderson_Fundamentals,
  editor    = {Douglas Henderson},
  publisher = {Marcel Dekker, Inc.},
  title     = {Fundamentals of Inhomogeneous Fluids},
  year      = {1992},
  address   = {New York},
  edition   = {1st},
  isbn      = {0-8247-8711-0},
}

@InBook{Lutsko2010,
  author    = {Lutsko, James F.},
  chapter   = {1},
  editor    = {Stuart A. Rice},
  pages     = {1-92},
  publisher = {Wiley},
  title     = {Recent Developments in Classical Density Functional Theory},
  year      = {2010},
  isbn      = {9780470564318},
  volume    = {144},
  booktitle = {Advances in Chemical Physics},
  doi       = {10.1002/9780470564318.ch1},
}

@Article{Henderson1991,
  author    = {Henderson, J.R.},
  journal   = {Mol. Phys.},
  title     = {A derivation of external fields that create step function density profiles at wall-fluid interfaces},
  year      = {1991},
  issn      = {1362-3028},
  month     = dec,
  number    = {5},
  pages     = {1125--1131},
  volume    = {74},
  doi       = {10.1080/00268979100102851},
  publisher = {Informa UK Limited},
}

@Article{Malijevsky2006,
  author    = {Malijevský, Alexandr},
  journal   = {J. Chem. Phys.},
  title     = {Alternative fundamental measure theory for additive hard sphere mixtures},
  year      = {2006},
  issn      = {1089-7690},
  month     = nov,
  number    = {19},
  volume    = {125},
  doi       = {10.1063/1.2393242},
  publisher = {AIP Publishing},
}

@Article{Roth2010,
  author    = {Roth, Roland},
  journal   = {J. Phys. Condens. Matter},
  title     = {Fundamental measure theory for hard-sphere mixtures: a review},
  year      = {2010},
  issn      = {1361-648X},
  month     = jan,
  number    = {6},
  pages     = {063102},
  volume    = {22},
  doi       = {10.1088/0953-8984/22/6/063102},
  publisher = {IOP Publishing},
}

@Article{Tarazona2000,
  author    = {Tarazona, P.},
  journal   = {Phys. Rev. Lett.},
  title     = {Density Functional for Hard Sphere Crystals: A Fundamental Measure Approach},
  year      = {2000},
  issn      = {1079-7114},
  month     = jan,
  number    = {4},
  pages     = {694--697},
  volume    = {84},
  doi       = {10.1103/physrevlett.84.694},
  publisher = {American Physical Society (APS)},
}

@Article{Cuesta2002,
  author    = {Cuesta, Jos\'e  A and Mart\'inez-Rat\'on, Yuri and Tarazona, Pedro},
  journal   = {J. Phys. Condens. Matter},
  title     = {Close to the edge of fundamental measure theory: a density functional for hard-sphere mixtures},
  year      = {2002},
  issn      = {0953-8984},
  month     = nov,
  number    = {46},
  pages     = {11965--11980},
  volume    = {14},
  doi       = {10.1088/0953-8984/14/46/307},
  publisher = {IOP Publishing},
}

@Article{Roth2002,
  author    = {Roth, R and Evans, R and Lang, A and Kahl, G},
  journal   = {J. Phys. Condens. Matter},
  title     = {Fundamental measure theory for hard-sphere mixtures revisited: the White Bear version},
  year      = {2002},
  issn      = {0953-8984},
  month     = nov,
  number    = {46},
  pages     = {12063--12078},
  volume    = {14},
  doi       = {10.1088/0953-8984/14/46/313},
  publisher = {IOP Publishing},
}

@Article{Vrugt2020,
  author    = {te Vrugt, Michael and Löwen, Hartmut and Wittkowski, Raphael},
  journal   = {Adv. Phys.},
  title     = {Classical dynamical density functional theory: from fundamentals to applications},
  year      = {2020},
  issn      = {1460-6976},
  month     = apr,
  number    = {2},
  pages     = {121--247},
  volume    = {69},
  doi       = {10.1080/00018732.2020.1854965},
  publisher = {Informa UK Limited},
}

\end{document}